\documentclass[a4paper,10pt]{article}
\usepackage{enumerate}
\usepackage{color}
\usepackage[utf8]{inputenc} 
\usepackage[english]{babel}
\usepackage[T1]{fontenc}
\usepackage{graphicx}
\usepackage{amsfonts,amssymb,amsmath,latexsym,amsthm}
\usepackage{textcomp}
\usepackage[pdftex]{hyperref}
\usepackage{geometry}
\DeclareMathOperator{\sech}{sech}

\title{Flow patterns induced by a moving disturbance in rotational flows within the forced Korteweg-de Vries equation}
\author{Marcelo V. Flamarion}
\date{}

\begin{document}
\maketitle
\begin{center}
{\footnotesize Unidade Acad{\^ e}mica do Cabo de Santo Agostinho, \\
UFRPE/Rural Federal University of Pernambuco, BR 101 Sul, Cabo de Santo Agostinho-PE, Brazil,  54503-900 \\
marcelo.flamarion@ufrpe.br }



\end{center}


\begin{abstract} 
Flow structures beneath a moving disturbance along a water free surface in the weakly nonlinear weakly dispersive
regime in a sheared channel with finite depth and constant vorticity are investigated. We compute the exact two branches of steady solutions in the disturbance moving frame. The velocity field in the bulk fluid is approximated which allows us to compute the flow structures beneath the free surface including stagnation points and Kelvin cat-eyes structures. We show that stagnation points  exist only in one branch of solutions. The bifurcation of the flow is analyzed according to the intensity of the vorticity and the speed of the moving disturbance. Differently from the unforced problem, stagnation points can arise for small values of the vorticity as long as the moving disturbance travels sufficiently fast.
	\end{abstract}

\section{Introduction}
The study of particle trajectories beneath  water waves traces its origins back to the seminal work of Stokes \cite{Stokes:1847} in 1847. Stokes demonstrated that under the conditions of irrotational flows and pure gravity waves, particle paths beneath nonlinear periodic waves form loops with minimal horizontal displacement in the direction of the wave propagation. This phenomenon, now recognized as Stokes drift, was investigated by him for both the shallow water and deep water regimes. Despite substantial interest in studying particle paths following Stokes work, it was only recently that rigorous results and theorems in this area were established by Constantin and his colleagues. Notably, Constantin and Villari \cite{Constantin:2008} demonstrated the absence of closed orbits in the linear scenario, and Constantin and Strauss \cite{Constantin:2010} revealed that closed orbits can always be found in the presence of a uniform counter-current. Subsequent numerical investigations of particle paths were conducted by Nachbin and Ribeiro-Jr \cite{Nachbin:2014}, utilizing a boundary integral method for the complete Euler equations. Additionally, asymptotic models like the Korteweg-de Vries equation (KdV) \cite{Johnson:1986, Kalisch:2012, Kalisch:2013, Gagnon:2017, Guan:2020}, Serre equations \cite{Khorsand:2014}, and Schrödinger equation \cite{Kalisch:2018, Kalisch:2020} have been employed to gain insights into particle trajectories within the context of the full Euler equations. One notable advantage of considering reduced models is their ability to provide accurate approximations to the complete nonlinear model with relatively low computational expenses. Moreover, in many instances, solutions can be derived in closed form, facilitating a more accessible theoretical perspective.

In flows with constant vorticity, an intriguing flow pattern known as the Kelvin cat-eye can emerge when the problem is framed within the context of the moving wave frame. This configuration is notable for the presence of stagnation points (critical points of the autonomous dynamical system governing particle trajectories) within the bulk fluid, accompanied by recirculation zones. For the scenario of periodic nonlinear waves, Teles Da Silva and Peregrine \cite{Teles:1988} employed a boundary integral formulation to capture enclosed streamlines in the full Euler equations framework. Later, anomalies in the pressure at the channel bottom were identified by Ribeiro-Jr et al. \cite{Ribeiro-Jr:2017} through the use of the conformal mapping method. Furthermore, investigating the linear Euler equations, Flamarion et al. \cite{Flamarion:2020} explored a time-dependent Kelvin cat-eye structure over a varying topography. In the case of solitary waves, Guan \cite{Guan:2020} computed recirculation zones (closed streamlines) beneath solitary waves for the Korteweg-de Vries equation, comparing the outcomes with those of the full nonlinear model. Guan demonstrated a strong agreement between particle trajectories from both models for solitary waves of small amplitudes. More recently, Flamarion \cite{WaveMotion} derived a fifth-order Korteweg-de Vries equation and delved into the flow patterns beneath generalized depression solitary waves. In this study, conditions leading to the emergence of stagnation points within the bulk fluid were investigated, along with the presentation of bifurcation diagrams.

In this work, we consider the  weakly nonlinear weakly dispersive regime to investigate particle paths beneath gravity waves produced by a moving disturbance along the free surface  in a sheared channel with constant vorticity. A forced Korteweg-de Vries equation  is derived asymptotically from the full Euler equations and the velocity field in the bulk fluid is obtained approximately. The problem is reformulated in the disturbance moving frame and stagnation points are found in the bulk fluid.  Furthermore, we show stagnation points appear in the bulk fluid even for small values of the vorticity parameter as long as the moving disturbance travels sufficiently fast. In addition, bifurcation diagrams in the space vorticity parameter vs. disturbance speed are discussed. In particular, we show that the flow can have zero, two or three stagnation points according to variations values of the vorticity parameter and the disturbance speed. This allows us to describe in great detail the recirculation zones attached to the bottom of the channel that emerge together with the stagnation points.

This article is organized as follows. In section 2 we present the mathematical formulation of the problem and derive a fKdV equation. In section 3 we obtain solitary wave solutions of the fKdV equation in closed form and present the results on particle paths. Section 4 is devoted to the conclusion. 
\section{Mathematical Formulation}
We examine a two-dimensional, incompressible flow of an inviscid fluid characterized by a constant density ($\rho$) within a channel of finite depth ($h$), influenced by a constant vorticity ($-\omega$) and acted upon by the force of gravity ($g$). Moreover, we introduce the presence of a moving disturbance ($\tilde{P}$) advancing at a constant speed ($U$) along the free surface ($\tilde{\eta}$). Within this context, the governing equations are the Euler equations
\begin{align} \label{eu1}
\begin{split}
& {\tilde{\phi}}_{xx}+{\tilde{\phi}}_{yy}= 0 \;\  \mbox{for} \;\ -h < y <{\tilde{\eta}}(x,t), \\
& {\tilde{\phi}}_{y} =0\;\ \mbox{at} \;\ y = -h, \\
& {\tilde{\eta}}_{t}+\eta_{x}\Big(\omega{\tilde{\eta}}+{\tilde{\phi}}_{x}\Big)-{\tilde{\phi}}_{y}=0
\;\ \mbox{at} \;\ y = {\tilde{\eta}}(x,t), \\
& {\tilde{\phi}}_{t}+\frac{1}{2}\Big({\tilde{\phi}}_{x}^2+2\omega{\tilde{\eta}}{\tilde{\phi}}_{x} +{\tilde{\phi}}_{y}^{2}\Big)+g{\tilde{\eta}}- \omega{\tilde{\psi}}= -\frac{\tilde{P}(x+Ut)}{\rho} \;\ \mbox{at} \;\ y = {\tilde{\eta}}(x,t).
\end{split}
\end{align}
Here, ${\tilde{\phi}}(x,y,t)$ represents the potential velocity, while ${\tilde{\psi}}(x,y,t)$ stands as its harmonic conjugate. For the sake of convenience, we find it advantageous to examine the system (\ref{eu1}) using dimensionless variables. To achieve this, let $\lambda$ denote a characteristic wavelength, $a$ a typical wave amplitude, and $c_0=(gh)^{1/2}$ denote the linear long-wave speed limit. Furthermore, introduce new variables  scaling as
\begin{align} \label{dimensionless}
\begin{split}
 & x\rightarrow\lambda x, \;\ y\rightarrow hy,  \;\ t\rightarrow \frac{\lambda}{c_0}t, \;\ \tilde{\eta}= a\eta, \\
 & \tilde{\phi}=\frac{ac_0\lambda}{h} \phi, \;\ \tilde{\psi}=ac_0\psi, \;\ \tilde{P}=\rho g h \frac{a^2}{h^2} P,
\end{split}
\end{align}
and sustituting (\ref{dimensionless}) in (\ref{eu1}) we obtain the dimensionless Euler equations
\begin{align} \label{eu2}
\begin{split}
& \mu^{2}{{\phi}}_{xx}+{{\phi}}_{yy}= 0 \;\  \mbox{for} \;\ -1 < y <{\epsilon{\eta}}(x,t), \\
& {\tilde{\phi}}_{y} =0\;\ \mbox{at} \;\ y = -1, \\
& {{\eta}}_{t}+\epsilon\eta_{x}\Big(\Omega{{\eta}}+{{\phi}}_{x}\Big)-\frac{1}{\mu^2}{{\phi}}_{y}=0
\;\ \mbox{at} \;\ y = {\epsilon{\eta}}(x,t), \\
& {\tilde{\phi}}_{t}+\frac{\epsilon}{2}\Big({{\phi}}_{x}^2+2\Omega{{\eta}}{{\phi}}_{x}+\frac{1}{\mu^2}{{\phi}}_{y}^{2}\Big) +{{\eta}}- \Omega{{\psi}}= -\epsilon P(x+Ft)  \;\ \mbox{at} \;\ y = {\epsilon{\eta}}(x,t).
\end{split}
\end{align}
In this context, $\mu=h/\lambda$ represents the shallow water parameter, $\epsilon=a/h$ characterizes the nonlinearity parameter, $-\Omega=-\omega h/c_0$ denotes the dimensionless vorticity parameter, and $F=U/(gh)^{1/2}$ is the Froude number. The Froude number quantifies the relationship between the velocity of the moving disturbance and the linear long-wave speed.

Our objective is to derive a fKdV equation through asymptotic analysis of equations (\ref{dimensionless}). With this aim, we focus on the weakly nonlinear regime ($\epsilon\approx 0$) coupled with weak dispersive effects ($\mu^2\approx 0$), where the nonlinearity and dispersion are balanced as $\epsilon=\mu$. Furthermore, we assume a power series expansion representation for the potential velocity, given by \cite{Whitham:1974}
\begin{equation}\label{phi}
\phi(x,y,t)=\sum_{n=0}^{\infty}f_{n}(x,t)(y+1)^{n}.
\end{equation}
Formally substituting  equation  (\ref{phi}) in condition  (\ref{eu2})$_{1,2}$, we obtain \begin{align} \label{exp0}
\begin{split}
& \phi(x,y,t) = \sum_{n=0}^{\infty}(-1)^{n}\frac{\epsilon^{n}}{(2n)!}\frac{\partial^{2n}\mathbf{\Phi}}{\partial x^{2n}}(y+1)^{2n}, \\
& \psi(x,y,t) = \sum_{n=0}^{\infty}(-1)^{n}\frac{\epsilon^{n}}{(2n+1)!}\frac{\partial^{2n+1}\mathbf{\Phi}}{\partial x^{2n+1}}(y+1)^{2n+1}, 
\end{split}
\end{align}
where $\mathbf{\Phi}(x,t)=\phi(x,-1,t)$ is the potential velocity evaluated at the bottom of the channel. Substituting equations (\ref{exp0})  into Kinematic and Bernoulli conditions (\ref{eu2})$_{2,3}$ and neglecting the second order terms as done by Guan \cite{Guan:2020} we obtain the Benney-Luke type equation
\begin{align} \label{exp}
\begin{split}
& \mathbf{\Phi}_{tt}-\mathbf{\Phi}_{xx}-\Omega\mathbf{\Phi}_{tx}-\epsilon\Big[\eta_{x}\mathbf{\Phi}_{x}+\Omega\eta\eta_x+\eta\mathbf{\Phi}_{xx}
\Big] \\
& -\epsilon\Big[\frac{1}{2}\mathbf{\Phi}_{ttxx}-\frac{1}{2}\big(\mathbf{\Phi}^{2}_{x}\big)_{t}-\frac{\Omega}{6}\mathbf{\Phi}_{txxx}-\frac{1}{6}\mathbf{\Phi}_{xxxx}\Big] - =-\epsilon F P_{x}(x+Ft).
\end{split}
\end{align}
Notice that substituting equations (\ref{exp0}) into (\ref{eu2}$)_4$, we have
\begin{equation}\label{neweta}
\eta = -\mathbf{\Phi}_{t}+\Omega\mathbf{\Phi}_{x}+\mathcal{O}(\epsilon).
\end{equation}
Substituing (\ref{neweta}) into equation (\ref{exp}) yields
\begin{align} \label{exp2}
\begin{split}
& \mathbf{\Phi}_{tt}-\mathbf{\Phi}_{xx}-\Omega\mathbf{\Phi}_{tx}-\epsilon\Big[\eta_{x}\mathbf{\Phi}_{x}+\Omega\eta\eta_x+\eta\mathbf{\Phi}_{xx}
\Big] \\
& -\epsilon\Big[\frac{1}{2}\mathbf{\Phi}_{ttxx}-\frac{1}{2}\big(\mathbf{\Phi}^{2}_{x}\big)_{t}-\frac{\Omega}{6}\mathbf{\Phi}_{txxx}-\frac{1}{6}\mathbf{\Phi}_{xxxx}\Big] =-\epsilon F P_{x}(x+Ft).
\end{split}
\end{align}

In order to derive a fKdV equation starting from equation (\ref{exp2}), we introduce traveling variables characterized by a ``slow time" component, defined as $\xi = x-ct$, and  $\tau=\epsilon t$, where $c$ is determined as the solution of the equation $c^{2}+\Omega c=1$ on the negative branch, namely
$$c = -\frac{\Omega}{2}-\frac{\sqrt{\Omega^2+4}}{2}.$$
Notice that with this choice, the speed $c$ is negative for all values of the vorticity parameter, consequently the Froude number can be chosen as
\begin{equation}\label{Froude}
F=-c+\epsilon f, 
\end{equation}
where $f$ is a parameter of order $\mathcal{O}(1)$. This parameter  defines how close the flow is to exact resonance ($F=-c$). In particular, the closest $f$ is to zero the stronger are the nonlinear effects.
Denote by $\mathbf{\Phi}(\xi,\tau)$ and $\eta(\xi,\tau)$ the potential velocity evaluated at the bottom of the channel and the free surface in the new coordinate system respectively. Thus, $\mathbf{\Phi}$ and $\eta$ satisfy the identities
\begin{align} \label{derivatives}
\begin{split}
& \mathbf{\Phi}_{x}=\mathbf{\Phi}_{\xi}, \\
& \mathbf{\Phi}_{t} =-c\mathbf{\Phi}_{\xi}+\epsilon\mathbf{\Phi}_{\tau}, \\
& \mathbf{\Phi}_{tt} =c^{2}\mathbf{\Phi}_{\xi\xi}-2c\epsilon\mathbf{\Phi}_{\xi\tau}+\mathcal{O}(\epsilon^{2}), \\
& \mathbf{\Phi}_{xt} =-c\mathbf{\Phi}_{\xi\xi}+\epsilon\mathbf{\Phi}_{\xi\tau}, \\
& {\eta} =(c+\Omega)\mathbf{\Phi}_{\xi}+\mathcal{O}(\epsilon).
\end{split}
\end{align}
Substituting  the relations (\ref{derivatives}) into (\ref{exp2}) and skipping some algebra we have
\begin{equation}\label{Phix}
\mathbf{\Phi}_{\xi\tau}+\frac{(\Omega^{2}+3)(c+\Omega)}{2c+\Omega}\mathbf{\Phi}_{\xi}\mathbf{\Phi}_{\xi\xi}+\frac{c^2}{3(2c+\Omega)}\mathbf{\Phi}_{\xi\xi\xi\xi}=-\frac{c}{2c+\Omega} P_{\xi}(\xi +f\tau).
\end{equation}
Lastly, substituting equation (\ref{derivatives})$_{5}$ into equation (\ref{Phix}) we have forced Korteweg-de Vries equation 
\begin{equation}\label{KdV}
\eta_{\tau}+\alpha\eta\mathbf{\eta}_{\xi}+\beta{\eta}_{\xi\xi\xi}=\gamma P_{\xi}(\xi +f\tau), 
\end{equation}
where the coefficients are given by
\begin{equation}\label{coefficients}
\alpha=\frac{\Omega^{2}+3}{2c+\Omega}, \;\ \beta=\frac{c^2}{3(2c+\Omega)} \mbox{ and } \gamma = \frac{-1}{2c+\Omega}.
\end{equation}
In particular, in the moving pressure frame the obtained fKdV (\ref{KdV}) reads
\begin{equation}\label{KdV2}
\eta_{\tau}+f\eta_{\xi}+\alpha\eta\mathbf{\eta}_{\xi}+\beta{\eta}_{\xi\xi\xi}= \gamma P_{\xi}(\xi).
\end{equation}
This equation governs the slow dynamics of the free surface when the Froude number ($F$) is close to $-c$.

\section{Results}
\subsection{Exact steady waves}

\begin{figure}[h!]
	\centering
	\includegraphics[scale=1]{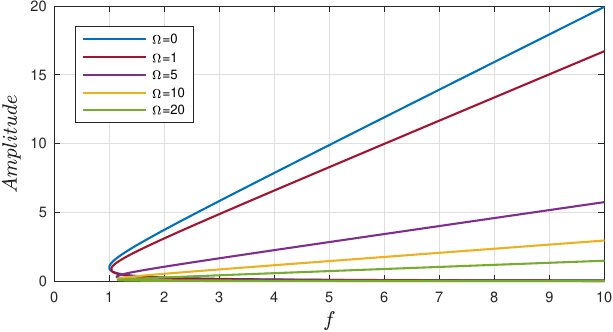} 
	\caption{The branches of steady solutions for different values of the vorticity parameter.}
	\label{Fig1}
\end{figure}

Exact solitary wave solutions of (\ref{KdV2}) can be achieved by appropriately selecting the disturbance profile through an inverse problem approach. Similar to the method employed by Chardard et al. \cite{Chardard}, we enforce
\begin{equation}
\eta(\xi) = A\sech^{2}(k \xi),
\end{equation}
to be a steady solution of equation (\ref{KdV2}). Thus, the disturbance satisfies 
$$\gamma P(\xi) = 2f\eta +\frac{\alpha}{2}{\eta}^{2}+\beta{\eta}_{\xi\xi}.$$
Consequently, we have that
\begin{equation}\label{Top}
P(\xi) = \frac{A}{2\gamma}\Bigg(\frac{2f+8\beta k^{2}}{\cosh^{2}(k \xi)}+\frac{A\alpha-12\beta k^2}{\cosh^{4}(k \xi)}\Bigg).
\end{equation}
Choosing the disturbance to be a $\sech^{2}$-type we obtain the two branch of solutions
\begin{align}\label{solution}
	\begin{split}
		& A = \frac{-\frac{3f}{\alpha}\pm \sqrt{\frac{9f^{2}}{\alpha^2}+\frac{12\gamma}{\alpha}G}}{2}, \\
		& k =\sqrt{\frac{A\alpha}{12\beta}},
	\end{split}
\end{align}
where $G$ is the amplitude of the disturbance. Notice that $\alpha<0$ and $\gamma/\alpha<0$ for all choices the vorticity parameter. Therefore, elevation solutions occur when $f\ge\sqrt{-(4/3) G \gamma\alpha}$. The branches of the amplitudes of the steady waves as a function of $f$ are depicted in Figure \ref{Fig1}. In both branches, the amplitude of the solitary wave decreases with the vorticity parameter. In the absence of vorticity ($\Omega=0$), Camassa and Wu \cite{Camassa1,Camassa2} showed that the perturbed solitary-wave solution (positive branch) is always unstable. On the other hand the perturbed uniform flow solution (negative branch) is stable only if $G\le\frac{80}{81}f^{2}$. In analogy to their works we say that the positive branch of (\ref{solution})  is the perturbed solitary-wave  solution and the negative branch is the perturbed  flow solution.

Once we are interested in examining particle trajectories for the complete model using the fKdV model as an approximation, the task at hand involves expressing both the free surface and the potential velocity in terms of Euler coordinates. The solitary wave, when represented in Euler coordinates, takes the form
\begin{equation}\label{solitaryEuler}
\eta(x,t)=A\sech^{2}\Big(k(x-(c-\epsilon f)t)\Big),
\end{equation}
and the  horizontal velocity evaluated at the bottom of the channel is
\begin{equation}\label{PotEuler}
\mathbf{\Phi}_{x}(x,t)= \frac{A}{c+\Omega}\sech^{2}\Big(k(x-(c-\epsilon f)t)\Big),
\end{equation}
where, $k$ and $A$ are defined in equations (\ref{solution}).

\subsection{Computing particle paths}
We compute particle paths beneath the solitary wave (\ref{solitaryEuler}) by solving the dynamical system 
\begin{align} \label{DSf}
\begin{split}
& \frac{dx}{dt} =\Omega y + \epsilon\phi_{x}(x,y,t)\approx \Omega y  + \epsilon\mathbf{\Phi}_{x}(x,t), \\
& \frac{dy}{dt} =\phi_{y}(x,y,t)\approx-\epsilon\mathbf{\Phi}_{xx}(x,t)(y+1). \\
\end{split}
\end{align}

For the purpose of computing stagnation points, we rewrite the system (\ref{DSf}) in the moving disturbance frame $X= x-(c-\epsilon f)t$ and $Y=y$. In this new framework, particle paths are solutions of the autonomous dynamical system
\begin{align} \label{DS}
\begin{split}
& \frac{dX}{dt} = \Omega Y -(c-\epsilon f) + \epsilon\mathbf{\Phi}_{X}(X), \\
& \frac{dY}{dt} =-\epsilon\mathbf{\Phi}_{XX}(X)(Y+1), \\
\end{split}
\end{align}
which are represented by the streamlines i.e.,  particle paths are the level curves of the stream function $\mathbf{\Psi}(X,Y)$, which is given by
\begin{equation} \label{stream function}
\mathbf{\Psi}(X,Y) = \epsilon\mathbf{\Phi}_{X}(X)(Y+1)+\frac{\Omega}{2}Y^{2}-(c-\epsilon f)Y.
\end{equation}

Flamarion et al. \cite{Marcelo-Paul-Andre} explored waves generated by the motion of a moving disturbance along the free surface using the Euler equations. They compared these findings with outcomes obtained from the fKdV equation in the regime of weak nonlinearity and weak dispersion, specifically for pressure distributions characterized by small amplitudes. Their investigation revealed that the solutions from both models exhibited strong agreement when the nonlinearity parameter was set to $\epsilon = 0.01$. Building upon these findings, in our present study, we maintain a consistent setting with $\epsilon = 0.01$ and $G = 1$ across all simulations presented. Thus, we anticipate a qualitative concurrence between the results regarding particle trajectories and those derived from the full model. It is worth noting that while there exist two branches of steady solitary waves, the emergence of stagnation points within the bulk fluid is restricted to the perturbed solitary-wave branch. As a result, we proceed by concentrating on solitary waves found in the positive branch of equation (\ref{solution}).
\begin{figure}[h!]
	\centering
	\includegraphics[scale=1]{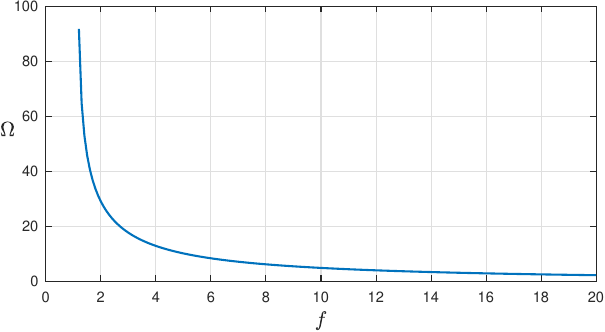} 
	\caption{The vorticity parameter as a function of $f$.}
	\label{Fig0}
\end{figure}
Stagnation points emerge initially at the channel bottom. Denoting the coordinates of the stagnation point below the crest of a solitary wave (\ref{solitaryEuler}) as $(0,Y^*)$, we find that the vertical position $Y^*$ satisfies the equation:
\begin{equation}\label{bottom}
0=\Omega Y^*(c-\epsilon f)+\epsilon\mathbf{\Phi}_{X}(0)=\Omega Y^*(c-\epsilon f)+\frac{\epsilon A}{c+\Omega}.
\end{equation}
The solutions to this equation depict a curve within the $\Omega\times f$ plane which is illustrated in Figure \ref{Fig0}. Notably, stagnation points materialize at the bottom of the channel, even when the vorticity parameter is small, given that the disturbance travels with sufficient speed. This stands in contrast to the unforced problem, where substantial vorticity values are needed for stagnation points to manifest. Below the curve, no stagnation points exist, while above it, a center emerges beneath the crest, along with two saddle points at the channel bottom. Denoting these two stagnation points as $(\pm X^*,-1)$ with $X^*>0$, their precise positions can be determined by solving the equation
\begin{equation}\label{saddles}
0=-\Omega -(c+\epsilon C)+\epsilon\mathbf{\Phi}_{X}(X^*)=-\Omega -(c+\epsilon C)+\frac{\epsilon A}{c+\Omega}\sech^{2}(kX^*)=0,
\end{equation}
for each specific pair $(f,\Omega)$ situated above the curve depicted in Figure \ref{Fig0}.
\begin{figure}[h!]
	\centering
	\includegraphics[scale=1.3]{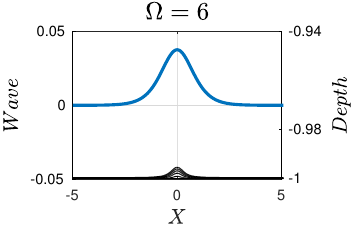} 
	\includegraphics[scale=1.3]{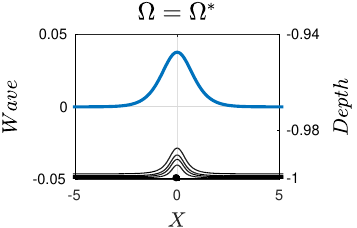} 
	\includegraphics[scale=1.3]{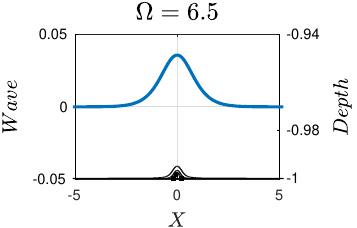} 
	\includegraphics[scale=1.3]{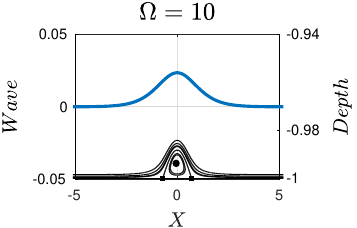} 
	\caption{Phase portraits  of (\ref{DS}) with $f=8$ and different values of the vorticity parameter.  Circles correspond to centres, squares to saddles at the bottom of the channel.}
	\label{Fig3}
\end{figure}

\begin{figure}[h!]
	\centering
	\includegraphics[scale=1.3]{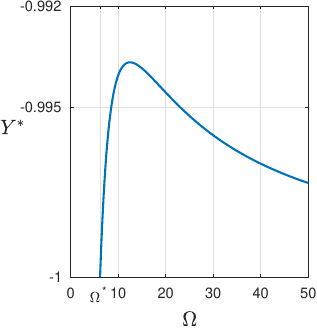} 
	\includegraphics[scale=1.3]{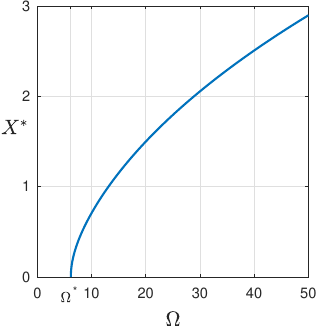}  
	\caption{Phase portraits  of (\ref{DS}) with $f=8$ and different values of the vorticity parameter.  Circles correspond to centres, squares to saddles at the bottom of the channel.}
	\label{Fig4}
\end{figure}

Figure \ref{Fig3} displays in great details the transition of the flow structure as the vorticity varies for $f=8$. For $\Omega=6$ there is no stagnation point in the fluid, as we increase the vorticity parameter to the critical value  $\Omega=\Omega^{*}\approx 6.1455$ a stagnation point appears at the bottom of the channel and then for $\Omega=6.5$ a centre appears in the bulk fluid and two saddles at the bottom forming a Kelvin cat-eye structure attached to the bottom of the channel. Notice that for  $\Omega=10$ a wider cat-eye structure arises. The exact positions of these stagnation points are shown in Figure \ref{Fig4}. The saddle $(X^*,-1)$ moves away from the origin as $\Omega$ increases while the centre $(0,Y^*)$ move upwards up to a certain value of the vorticity parameter and then move downwards.

The presented results are expect to agree qualitatively well with the full nonlinear model. The benefit of employing the asymptotic framework in this context is underscored by its ability to efficiently calculate stagnation points, bifurcation diagrams, and intricate phenomena like Kelvin cat-eyes. Exploring a comparison between the outcomes obtained in this study and those derived from the Euler equations appears to be a logical direction for future investigation.

\section{Conclusions}
In this paper, we have investigated the flow structures beneath a moving disturbance in the presence of a vertically sheared current with constant vorticity using the fKdV equation. We  determined two exact branches solutions for the fKdV equation in the moving disturbance frame. We showed that stagnation points arise in the bulk fluid only in one of the branches even for small values of the vorticity parameter as long as the moving disturbance moves fast enough.

\section{Acknowledgements}
M.V.F is grateful to IMPA for hosting him as visitor during the 2023 Post-Doctoral Summer Program.

	\section*{Declarations}
	
	\subsection*{Conflict of interest}
	The authors state that there is no conflict of interest. 
	\subsection*{Data availability}
	
	Data sharing is not applicable to this article as all parameters used in the numerical experiments are informed in this paper.

\end{document}